\RequirePackage[T1]{fontenc}
\documentclass[10pt,journal]{IEEEtran}

\usepackage{amsmath,amssymb,amsthm,bm,booktabs,cite}
\usepackage[caption=false]{subfig}
\usepackage{titlesec}
\titlespacing*{\section}
  {0pt}{1.0ex plus 0.2ex minus 0.2ex}{0.5ex}
\titlespacing*{\subsection}
  {0pt}{0.8ex plus 0.2ex minus 0.2ex}{0.4ex}
\newtheorem{theorem}{Theorem}
\newtheorem{proposition}{Proposition}

\newcommand{\thv}{\bm\theta}
\newcommand{\HH}{\bm H}
\newcommand{\WW}{\bm W}
\newcommand{\GG}{\bm G}
\newcommand{\RR}{\mathbb R}
\DeclareMathOperator{\diag}{diag}
\DeclareMathOperator{\Tr}{Tr}

\usepackage{graphicx}
\begin{document}
\title{Sharp Minimum-Rate Guarantees for\protect\\ Power-Based Discrete RIS Design}
\author{Hernan~X.~C{\'o}rdova\thanks{H. X. C{\'o}rdova is with the Electrical Engineering Department, Signal Processing Systems, Eindhoven University of Technology, The Netherlands (e-mail: h.x.cordova@tue.nl).}\thanks{This work will be submitted to the IEEE for possible publication. Copyright may be transferred without notice, after which this version may no longer be accessible.}}
\markboth{Preprint, to be submitted to IEEE Wireless Communications Letters}{C{\'o}rdova: Sharp Minimum-Rate Guarantees for Power-Based Discrete RIS Design}
\maketitle
\begin{abstract}
We characterise when desired-power optimisation guarantees the minimum user rate (MUR) in discrete reconfigurable intelligent surface systems with fixed precoding. For the weighted sum of desired powers, a bounded channel exists whose exact maximiser starves a user: no positive guarantee exists. For the minimum desired power, the retained fraction of the optimal MUR is at least a sharp constant set by the phase's power quality and residual interference. A semidefinite dual certifies it without enumerating phases. Minimum-power phases keep $88\%$ of the optimum under weak reflection and about half under strong reflection; omitted interference causes most of the loss.
\end{abstract}
\section{Introduction}
\begin{figure}[t]
\centering
\includegraphics{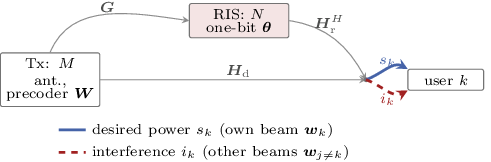}
\caption{Signal model. The one-bit phase $\thv$ sets the channel $\HH(\thv)=\HH_{\rm d}+\HH_{\rm r}^H\diag(\thv)\GG$. Through it, user $k$ gets a desired signal from its own beam $\bm w_k$ (power $s_k$) and interference from the other beams $\bm w_{j\ne k}$ (power $i_k$); both move with $\thv$. A desired-power objective counts only $s_k$, while the rate uses $s_k/(1+i_k)$.}
\label{fig:system}
\end{figure}

\IEEEPARstart{R}{econfigurable} intelligent surfaces (RIS) steer a multiuser downlink by setting the phase of each passive element~\cite{direnzo2020smart,wu2019irs,basar2019ris}. A common and convenient choice is to pick the phases that maximise a desired-power objective, which is quadratic in the phases and suits exact, approximate, and quantum solvers~\cite{pekcan2024power,paul2026risqaoa,gulbahar2026risqaoa,krikidis2025risqa}. With independent streams, the weakest user's rate depends not only on its desired power but also on the interference the same phases create (Fig.~\ref{fig:system}), which a desired-power objective ignores.

Two families of designs bound the resulting rate. Rate-directed designs keep interference explicit, through individual signal-to-interference-plus-noise ratio (SINR) constraints~\cite{schubert2004mmf,wu2020discrete}, achievable-rate optimisation~\cite{di2020hybrid}, relaxed max--min SINR targets~\cite{xie2021fairness}, deterministic equivalents~\cite{nadeem2020maxmin}, or joint cell-free beamforming~\cite{jin2023cellfree}. Power-directed designs, including the quantum searches above, keep only desired power. The two coincide only in single-group multicast, where a shared stream removes interference~\cite{yan2023multicast,sidiropoulos2006multicast}; multigroup bounds then handle intergroup interference~\cite{chang2008bounds}.

For independent streams and a fixed precoder, these families optimise different objectives, and how much minimum user rate (MUR) a power-directed design keeps is not known. This letter settles that question. We fix the precoder, and for a phase chosen by a power objective we ask what fraction of the optimal MUR it is guaranteed to retain. The answer separates the two power objectives sharply:
\begin{itemize}\setlength{\itemsep}{1pt}\setlength{\parsep}{0pt}\setlength{\topsep}{2pt}\setlength{\partopsep}{0pt}
\item[(i)] The weighted sum of desired powers gives no guarantee: on a bounded single-element channel its exact maximiser drives one user's rate to zero while the interference stays bounded, and one such channel rules out any universal factor (Prop.~\ref{prop:starvation}, App.~\ref{app:prop1}).
\item[(ii)] The minimum desired power (MDP) gives a tight guarantee: a phase that nearly maximises the MDP keeps at least a fixed fraction of the optimal MUR, set by how close it is to the best desired power and by the interference it leaves. No larger fraction is valid---two-element channels attain it with equality, and it is approached even when the power optimum is unique---and the guarantee holds at every SINR, weakest when the desired signal is weak (Thm.~\ref{thm:transfer}, App.~\ref{app:thm1}).
\item[(iii)] A semidefinite dual turns the guarantee into a certificate that bounds the retained fraction for a given phase without enumerating phases; an exact decomposition splits its slack into a power-bound part and an interference part (Prop.~\ref{prop:dual}, App.~\ref{app:prop2},~\eqref{eq:gap}).
\end{itemize}
The analysis uses only standard inequalities and a semidefinite relaxation. Sec.~\ref{sec:model} fixes the model and objectives; Sec.~\ref{sec:theory} proves the guarantees and the dual certificate; and Sec.~\ref{sec:num} confirms them by exhaustive optimisation, showing the shortfall persists across precoders, array sizes, transmit correlation, and phase resolution.

\noindent\emph{Notation.} Bold lower- and upper-case letters are vectors and matrices. $(\cdot)^T$, $(\cdot)^H$, $\Re\{\cdot\}$, $\diag(\cdot)$, and $\Tr(\cdot)$ denote transpose, conjugate transpose, real part, diagonal matrix, and trace; $A{\succeq}0$ is positive semidefinite, $I_n$ the $n{\times}n$ identity, and a star an exact optimum over $\Theta$.
\section{Model and Objectives}\label{sec:model}
Consider $M$ transmit antennas, $K$ single-antenna users, and $N$ passive RIS elements with one-bit phases $\thv\in\Theta=\{-1,+1\}^N$. The composite channel is:
\begin{equation}
 \HH(\thv)=\HH_{\rm d}+\HH_{\rm r}^{H}\diag(\thv)\GG,
 \label{eq:channel}
\end{equation}
where $\HH_{\rm d}\in\mathbb C^{K\times M}$, $\HH_{\rm r}\in\mathbb C^{N\times K}$, and $\GG\in\mathbb C^{N\times M}$ are the direct, RIS--user, and transmitter--RIS channels, so each of the two additive terms in~\eqref{eq:channel} is $K\times M$. The precoder $\WW=[\bm w_1,\ldots,\bm w_K]$ is fixed and satisfies a total-power budget $\Tr(\WW\WW^H)\le P$. Channel coefficients are finite and the noise powers obey $\sigma_k^2>0$. The guarantees of Sec.~\ref{sec:theory} are deterministic and per-channel, with no distributional assumption; Sec.~\ref{sec:num} stress-tests them on specific deployments, including transmit-correlated ones.

Let $\bm h_k^H(\thv)$ denote row $k$ of $\HH(\thv)$. The noise-normalised desired and interference powers of user $k$ are:
\begin{equation}
 s_k=\frac{|\bm h_k^H\bm w_k|^2}{\sigma_k^2},\qquad
 i_k=\frac{\sum_{j\ne k}|\bm h_k^H\bm w_j|^2}{\sigma_k^2}.
 \label{eq:powers}
\end{equation}
We suppress the phase argument when it is clear. The minimum SINR and the MUR are:
\begin{equation}
 \begin{aligned}
 g(\thv)&=\min_k\frac{s_k(\thv)}{1+i_k(\thv)},\\
 R(\thv)&=\log_2(1+g(\thv)),\qquad R^\star=\max_{\Theta}R.
 \end{aligned}
 \label{eq:rate}
\end{equation}
We compare two power objectives:
\begin{equation}
 P_\omega(\thv)=\sum_k\omega_k s_k(\thv),\qquad
 D(\thv)=\min_k s_k(\thv),
 \label{eq:objectives}
\end{equation}
with fixed weights $\omega_k\ge0$, $\sum_k\omega_k=1$, and $D^\star=\max_\Theta D$, $g^\star=\max_\Theta g$. The weighted sum is a quadratic unconstrained binary optimisation (QUBO), and $D$ a minimum of quadratics. Both ignore interference. We assume $R^\star>0$ (hence $g^\star>0$ and $D^\star>0$). The streams are independent, proper complex Gaussian, and unit power, each decoded treating interference as noise.

\noindent\emph{Problem Objective.} We bound $R(\hat\thv)/R^\star$ for a phase $\hat\thv$ chosen by $P_\omega$ or $D$, and identify when no positive bound exists. The fixed precoder sets this apart from joint max--min precoder design~\cite{cordova2026cl}.
\section{Power-to-Rate Guarantees}\label{sec:theory}
\subsection{Weighted Power Has No Guarantee}
\begin{proposition}[Weighted power can starve a user]
\label{prop:starvation}
Fix weights $\omega_k{\ge}0$ with $\sum_k\omega_k{=}1$, and let $\hat\thv_\omega\in\arg\max_{\thv\in\Theta}P_\omega(\thv)$ be any exact weighted-power maximiser. There is a one-parameter family of channels of the form~\eqref{eq:channel}, with $K{=}2$ users, $M{=}2$ antennas, $N{=}1$ element, bounded coefficients and interference-to-noise ratios $i_k{\le}1$, indexed by $\epsilon\in(0,\tfrac12]$, on which $R(\hat\thv_\omega)/R^\star\to0$ as $\epsilon\downarrow0$. Hence $\inf_\epsilon R(\hat\thv_\omega)/R^\star=0$: no constant $c{>}0$ gives $R(\hat\thv_\omega)\ge cR^\star$ for all channels.
\end{proposition}
\begin{proof} See App.~\ref{app:prop1}. \end{proof}
In the construction of App.~\ref{app:prop1}, as $\epsilon\downarrow0$ the weighted sum $P_\omega$ rises while user~2's desired power $s_2{=}\epsilon^2$ vanishes, so user~2's rate, and with it the MUR, goes to zero. Enforcing equal rates can lower the total throughput, the price of fairness~\cite{bertsimas2011}; the construction shows more, that $R(\hat\thv_\omega)/R^\star\to0$, so \emph{no} positive fraction of the MUR survives. A single bounded channel with $i_k{\le}1$ already rules out any universal factor, so capping the interference cannot justify replacing the minimum rate by the weighted power.
\subsection{A Sharp Guarantee for Minimum Desired Power}
\begin{theorem}[Sharp power-to-rate guarantee]
\label{thm:transfer}
For any candidate $\hat\thv\in\Theta$, suppose $D(\hat\thv)\ge\alpha D^\star$, where $0\le\alpha\le1$. Define $\hat\kappa=\max_k i_k(\hat\thv)$ and $c=\alpha/(1+\hat\kappa)$. Then:
\begin{equation}
 R(\hat\thv)\ge\log_2\!\left(1+c(2^{R^\star}-1)\right)
 \ge cR^\star.
 \label{eq:transfer}
\end{equation}
The constant $c$ cannot be replaced by any larger one: for each admissible $\alpha$ and $\hat\kappa$ there are bounded two-element channels on which~\eqref{eq:transfer} holds with equality, and the bound is still approached when the phase that maximises $D$ is unique.
\end{theorem}
The guarantee follows from the inequalities in App.~\ref{app:thm1}, and its sharpness from a matching two-element construction there. The linear bound $cR^\star$ is the worst case over channels whose desired signal can be made arbitrarily weak. When $0<c<1$ and $R^\star>0$, the nonlinear bound is strictly stronger.

The bound uses the interference only at the selected phase. Its linear factor is weakest when the desired signal is weak: for $\hat\kappa=1$ the retained fraction approaches $1/2$ (App.~\ref{app:thm1}). It also separates objective from solver, since maximising $P_\omega$ need not meet a target $\alpha$ for $D$. The only deployment-dependent quantity in $c$ is $\hat\kappa$, so the guarantee is weak wherever interference is large.
\subsection{A Computable Certificate}
The premise must be checkable. For one-bit phases, the lifted vector $\bm x=(1,\thv^T)^T\in\{-1,+1\}^{N+1}$ is real. Writing the desired gain $\bm h_k^H\bm w_k/\sigma_k$ as $\bm v_k^T\bm x$ gives $s_k=|\bm v_k^T\bm x|^2=\bm x^T Q_k\bm x$, with $Q_k=\Re\{\overline{\bm v}_k\bm v_k^T\}$ real symmetric and positive semidefinite on real vectors. Relaxing the rank-one requirement $X=\bm x\bm x^T$ to $X\succeq0$ gives the standard semidefinite program (SDP)~\cite{goemans1995maxcut,luo2010sdr}:
\begin{equation}
 \max_{X,t}\ t\quad\text{s.t.}\quad
 \Tr(Q_kX)\ge t\ (\forall k),\quad \diag X=\bm1,\quad X\succeq0.
 \label{eq:sdp}
\end{equation}
Relaxing enlarges the feasible set, so the optimum of~\eqref{eq:sdp} is at least $D^\star$ and is thus an upper bound on it. By weak duality, any dual-feasible point gives such an upper bound without solving~\eqref{eq:sdp} exactly.
\begin{proposition}[Dual rate certificate]
\label{prop:dual}
Let $\lambda_k\ge0$, $\lambda_\Sigma=\sum_k\lambda_k>0$, and $\bm y\in\RR^{N+1}$ satisfy:
\begin{equation}
 A=\diag(\bm y)-\sum_k\lambda_kQ_k\succeq0,
 \qquad U=\frac{\bm1^T\bm y}{\lambda_\Sigma}>0.
 \label{eq:dual}
\end{equation}
Then any feasible candidate satisfies:
\begin{equation}
 \frac{R(\hat\thv)}{R^\star}\ge
 C(\hat\thv):=\frac{R(\hat\thv)}{\log_2(1+U)}
 \ge\frac{D(\hat\thv)}{U(1+\hat\kappa)}.
 \label{eq:certificate}
\end{equation}
\end{proposition}
\begin{proof} See App.~\ref{app:prop2}. \end{proof}
Thus $D(\hat\thv)/U$ is a certified $\alpha$, and $C$ sharpens it with the measured SINR; we obtain the dual with Clarabel~\cite{clarabel2026}.

A tighter comparison is the classical SINR-target test~\cite{schubert2004mmf,xie2021fairness}, which keeps the interference $i_k=\bm x^TB_k\bm x$, with $B_k\succeq0$ built from the interfering amplitudes as $Q_k$ is built from the desired one: applying~\eqref{eq:sdp} with $Q_k-\tau B_k$ at a trial SINR $\tau$, a dual bound below $\tau$ proves no phase reaches $g\ge\tau$, so $R(\hat\thv)/\log_2(1+\tau)$ is a certificate (Fig.~\ref{fig:certificate}).

To interpret the certificate looseness, let $R_U=\log_2(1+U)$ and $R_D=\log_2(1+D^\star)$. The exact diagnostic identity is:
\begin{equation}
 R_U-R^\star=
 \underbrace{\log_2\frac{1+U}{1+D^\star}}_{\text{power upper-bound gap}}+
 \underbrace{\log_2\frac{1+D^\star}{1+g^\star}}_{\text{interference-omission gap}}.
 \label{eq:gap}
\end{equation}
Both terms are nonnegative; computing~\eqref{eq:gap} needs the exact optima, so it is a diagnostic only.

\section{Numerical Evaluation}\label{sec:num}
We take $K{=}3$, $M{=}8$, $N{=}12$, $\sigma_k^2{=}0.1$, with direct and RIS--user coefficients independent circular complex Gaussian of unit variance and $\GG$ of variance $1/N$. A factor $\rho\in\{0.3,1,2\}$ on the RIS--user coefficients sets the reflected-path strength, and each setting has $150$ independent channel realisations. The precoder is the normalised regularised zero-forcing $\WW\propto\HH_{\rm d}^H(\HH_{\rm d}\HH_{\rm d}^H+0.1\,I_K)^{-1}$, fixed across phases and scaled to unit total power $\|\WW\|_F{=}1$. Across these realisations the all-$+1$ phase is never optimal, so the phase choice matters.

Four controls each use $257$ phase evaluations: uniform sampling with rate selection, and restarted steepest one-bit searches directed by rate, mean or minimum desired power, averaged over $16$ repetitions per channel. Exhaustive enumeration of all $4\,096$ phases supplies $R^\star$, and $5\,000$-resample channel bootstraps give the intervals.
\begin{table}[t]
\centering
\caption{Cohort mean achieved rate fraction $R/R^\star$ ($150$ realisations/setting).}
\label{tab:controls}
\footnotesize
\setlength{\tabcolsep}{4pt}
\begin{tabular}{@{}lccc@{}}
\toprule
Selection objective / search&$\rho{=}0.3$&$\rho{=}1$&$\rho{=}2$\\
\cmidrule(r){1-1}\cmidrule(l){2-4}
Rate / uniform samples&0.973&0.890&0.823\\
Rate / one-bit search&0.992&0.967&0.949\\
Mean power / one-bit search&0.863&0.573&0.413\\
Minimum power / one-bit search&0.876&0.611&0.551\\
\addlinespace
Mean power / exhaustive&0.863&0.574&0.411\\
Minimum power / exhaustive&0.878&0.611&0.558\\
\bottomrule
\end{tabular}
\end{table}
\begin{figure}[t]
\centering
\includegraphics{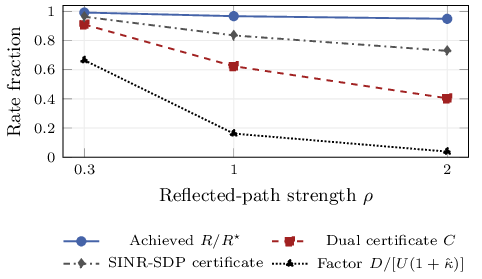}
\vspace{-0.1cm}
\caption{Rate-directed one-bit search: achieved and certified rate fractions ($150$ realisations, $16$ repetitions each). The novelty is the certificate: the dual $C$ and SINR-target curves stay below the achieved fraction at every $\rho$, so they certify it without enumerating phases. All $95\%$ channel-bootstrap intervals have half-width below $0.014$ and lie within the plot markers.}
\label{fig:certificate}
\end{figure}
Table~\ref{tab:controls} reports the mean retained fraction. Rate-directed search stays near the optimum, while both power-directed rules fall well below it, and exhaustive power optimisation performs within the bootstrap interval of the budgeted search. The minimum-power rule has the highest certified power quality yet a much lower rate, because it leaves several times the interference of the rate-directed rules. That maximising desired power need not maximise rate is well understood and underlies rate-directed designs~\cite{schubert2004mmf,xie2021fairness}; new here is the \emph{size} of the shortfall, which~\eqref{eq:gap} attributes to omitted interference. Since the exhaustive and budgeted optima agree, the loss is the \emph{objective mismatch}, not an imperfect search.

Fig.~\ref{fig:certificate} plots the achieved fraction against the three certificates of Sec.~\ref{sec:theory}: the dual $C$ and the SINR-target certificate stay below it at every $\rho$, and the linear factor $D/[U(1+\hat\kappa)]$ is the loosest.
\begin{table}[t]
\centering
\caption{Mean gaps to the rate upper bound, in bit/s/Hz ($150$ realisations/setting). The percentage is a ratio of cohort means; per-channel mean shares are $89.3$, $93.1$, $93.4\%$.}
\label{tab:gaps}
\footnotesize
\begin{tabular}{lccc}
\toprule
Gap component&$\rho{=}0.3$&$\rho{=}1$&$\rho{=}2$\\
\midrule
Power upper bound: $R_U-R_D$&0.037&0.123&0.229\\
Interference omission: $R_D-R^\star$&0.336&1.725&3.274\\
Total: $R_U-R^\star$&0.373&1.848&3.503\\
\addlinespace
Interference share of total [\%]&90.2&93.3&93.5\\
\bottomrule
\end{tabular}
\end{table}
Table~\ref{tab:gaps} applies~\eqref{eq:gap} to the exact optima. Omitted interference is the larger term at every reflection strength, $90\%$ to $94\%$ of the total by cohort mean, exceeding the power-bound term on $449$ of $450$ realisations ($99.8\%$). Both power optima are unique on every realisation, and every dual is repaired to exact feasibility, so each upper bound is valid.

\begin{figure}[t]
\centering
\includegraphics{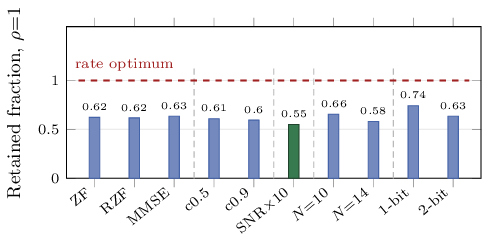}
\vspace{-0.15cm}
\caption{Minimum-power retained fraction at $\rho{=}1$ (same $\rho$ as Fig.~\ref{fig:certificate}) under single changes to the baseline ($K{=}3$, $M{=}8$, $N{=}12$ one-bit elements, regularised zero-forcing; second bar). Groups, left to right: the three precoders; transmit correlation $0.5$ and $0.9$; a tenfold SNR rise (green); element count $N{=}10,14$; and phase resolution. The 1-bit and 2-bit bars are computed at $N{=}6$, where the $4^6{=}4{,}096$-phase two-bit alphabet matches the $2^{12}$ one-bit enumeration budget; the drop from $0.74$ to $0.63$ is the within-$N{=}6$ effect of finer resolution. The dashed line is the rate optimum (fraction $1$).}
\label{fig:bars}
\end{figure}

\noindent\emph{Robustness.} Theorem~\ref{thm:transfer} makes the shortfall depend on the deployment only through the interference $\hat\kappa$ at the selected phase, so a change that leaves interference present cannot remove it. Fig.~\ref{fig:bars} bears this out at $\rho{=}1$ over $400$ realisations per configuration (more than the $150$ above, so the ten configurations separate). The baseline is minimum-power selection with regularised zero-forcing at $N{=}12$ (second bar). The retained fraction stays well below the optimum under every single change. It barely moves across the three precoders (ZF, RZF, MMSE agree to within $0.02$, as all leave comparable residual interference) or under correlation; it drops for larger arrays and finer resolution; and it is lowest at a tenfold SNR rise (green, $0.55$), where the link is interference-limited and ignoring interference costs most. A paired sign-rank test rejects equality of the power-selected and rate-optimal minimum rates in every setting ($p<10^{-4}$). A sweep to $M{=}512$ antennas and $K{=}20$ users keeps the fraction below $0.9$ and $\hat\kappa$ above $10$ throughout.
\section{Conclusion}\label{sec:conclusion}
The minimum desired power gives a tight rate guarantee, set by the interference at the chosen phase, while the weighted sum gives none. A dual certifies it for a given phase, and an exact identity shows that omitted interference, not the power bound, causes most of the loss ($90\%$ to $94\%$); exhaustive search does not close it. A RIS design should therefore certify the selected phase's rate rather than read fairness from its power. The gap persists across precoders and correlation, widening at higher SNR and (at $N{=}6$) finer resolution, and persisting below $0.9$ up to $512$ antennas and $20$ users.
\appendices
\section{Proof of Proposition~\ref{prop:starvation}}
\label{app:prop1}
Given the weights, relabel the users so that $\omega_1=\max_k\omega_k\ge\tfrac12$ (possible since $K{=}2$), and let user~1 be the reinforced one. Set $\WW=I_2/\sqrt2$, $\sigma_k^2=1/2$, $\bm u_\epsilon=(1,-1+\epsilon)^T$, $\bm e=(1,1)^T$ with $0\le\epsilon\le\tfrac12$, and $\HH(\theta)=I_2+\tfrac{1-\theta}{2}\bm u_\epsilon\bm e^T$, a bounded instance of~\eqref{eq:channel} with $(\HH_{\rm d},\HH_{\rm r}^{H},\GG)=(I_2{+}\tfrac12\bm u_\epsilon\bm e^T,{-}\bm u_\epsilon,\tfrac12\bm e^T)$. The two phases give:
\begin{equation}
\begin{aligned}
\theta{=}{+}1:&\quad \HH=I_2,\ \ (s_k,i_k)=(1,0),\ \ R(+1)=1,\\
\theta{=}{-}1:&\quad (s_1,s_2)=(4,\epsilon^2),\ \ (i_1,i_2)=(1,(1{-}\epsilon)^2),
\end{aligned}
\label{eq:a-phases}
\end{equation}
so $R^\star{=}R(+1){=}1$ and $\max_{k,\theta}i_k{=}1$, while at $\theta{=}{-}1$ user~2 starves:
\begin{equation}
\frac{R(-1)}{R^\star}=\log_2\!\Big(1+\tfrac{\epsilon^2}{1+(1-\epsilon)^2}\Big)\ \longrightarrow\ 0 \quad(\epsilon\downarrow0).
\label{eq:a-limit}
\end{equation}
This phase is selected for \emph{every} weighting: since $\omega_1{\ge}\tfrac12$, thus:
\begin{equation}
P_\omega(-1){=}4\omega_1+\epsilon^2\omega_2\ {\ge}\ 4\omega_1\ {\ge}\ 2\ {>}\ 1{=}P_\omega(+1),
\label{eq:a-argmax}
\end{equation}
so $\hat\thv_\omega{=}{-}1$ is the unique maximiser and $R(\hat\thv_\omega)/R^\star{\to}0$. Finally, rescaling $\HH$ by $\sqrt c$ sends $i_k{\mapsto} c\,i_k$ while keeping $R^\star{\ge}\log_2(1+c)>0$, so no interference cap $i_k{\le} c$ removes the obstruction.\hfill$\blacksquare$
\section{Proof of Theorem~\ref{thm:transfer}}
\label{app:thm1}
Each $s_k/(1+i_k)\le s_k$ gives $g\le D$ pointwise, hence $g^\star\le D^\star$. Bounding $i_k{\le}\hat\kappa$ and using $D(\hat\thv)\ge\alpha D^\star\ge\alpha g^\star$, leads to:
\begin{equation}
g(\hat\thv)\ge\frac{D(\hat\thv)}{1+\hat\kappa}\ge\frac{\alpha D^\star}{1+\hat\kappa}\ge c\,g^\star,\qquad c=\frac{\alpha}{1+\hat\kappa}.
\label{eq:chain}
\end{equation}
With $g^\star{=}2^{R^\star}{-}1$, monotonicity of $\log_2(1+\cdot)$ turns~\eqref{eq:chain} into the first bound in~\eqref{eq:transfer}; concavity through the origin, $\log_2(1+cx)\ge c\log_2(1+x)$ for $x{\ge}0$ and $c{\in}[0,1]$, then gives the linear bound $cR^\star$.

\noindent
For sharpness, take $K{=}M{=}N{=}2$, $\WW{=}I_2/\sqrt2$, $\sigma_k^2{=}1/2$, so that $s_k{=}|H_{kk}|^2$ and $i_k{=}|H_{kj}|^2$, with:
\begin{equation}
\HH(\thv)=\begin{bmatrix} a+\delta\tfrac{1-\theta_1}{2} & \sqrt\kappa\tfrac{1-\theta_1}{2}\\[2pt] \sqrt\kappa\tfrac{1-\theta_2}{2} & a+\delta\tfrac{1-\theta_2}{2}\end{bmatrix}.
\label{eq:a-sharpH}
\end{equation}
Setting $\delta{=}(\sqrt\alpha-1)a$ makes:
\begin{equation}
\begin{aligned}
(+1,+1):&\quad D^\star=g^\star=a^2,\\
\hat\thv{=}({-}1,{-}1):&\quad D=\alpha a^2,\ \hat\kappa=\kappa,\ g=\tfrac{\alpha a^2}{1+\kappa}=c\,g^\star,
\end{aligned}
\label{eq:a-sharpval}
\end{equation}
so~\eqref{eq:transfer} holds with equality, and $R(\hat\thv)/R^\star\to\alpha/(1+\kappa)$ as $a\downarrow0$. For the $\alpha{=}1$ case set instead $\delta=a^2$: the $D$-maximiser moves to $({-}1,{-}1)$ with $D^\star{=}a^2(1+a)^2$, while the $g$-maximiser stays at $(+1,+1)$ with $g^\star{=}a^2$; if $(1{+}a)^2{<}1{+}\kappa$ both are unique and distinct, and $R(\hat\thv)/R^\star{\to}1/(1{+}\kappa)$. Thus $c$ is approached even when the $D$-maximiser is unique.\hfill$\blacksquare$
\section{Proof of Proposition~\ref{prop:dual}}
\label{app:prop2}
Lift to $\bm x{=}(1,\thv^T)^T$, so $x_j^2{=}1$ and $s_k{=}\bm x^TQ_k\bm x{=}|\bm v_k^T\bm x|^2{\ge}0$ (the antisymmetric part of $\overline{\bm v}_k\bm v_k^T$ vanishes in a real quadratic form). Since $A\succeq0$, then:
\begin{equation}
0{\le}\bm x^TA\bm x{=}\textstyle\sum_j y_j x_j^2{-}\sum_k\lambda_k s_k=\bm1^T\bm y-\sum_k\lambda_k s_k,
\label{eq:c-psd}
\end{equation}
so dividing by $\lambda_\Sigma$ gives $D(\thv){=}\min_k s_k{\le}\tfrac1{\lambda_\Sigma}\sum_k\lambda_k s_k{\le} U$ for every $\thv$. Hence $g^\star{\le} D^\star{\le} U$, so $R^\star{\le}\log_2(1{+}U)$, and dividing the nonnegative $R(\hat\thv)$ by the larger denominator gives $R(\hat\thv)/R^\star{\ge} R(\hat\thv)/\log_2(1{+}U){=}C$. Finally, with $q=D(\hat\thv)/(U(1+\hat\kappa)){\in}[0,1]$, the chain~\eqref{eq:chain} gives $qU{\le} g(\hat\thv)$, so:
\begin{equation}
R(\hat\thv)\ge\log_2(1+qU)\ge q\log_2(1+U),
\label{eq:c-chord}
\end{equation}
i.e. $C\ge q=D(\hat\thv)/(U(1+\hat\kappa))$.\hfill$\blacksquare$

{\footnotesize
\setlength{\IEEEilabelindent}{0pt}
\bibliographystyle{IEEEtran}
\bibliography{references_final_WCLarxiv}}
\end{document}